\documentclass[a4paper,11pt]{article}
\pdfoutput=1
\usepackage[noblocks]{authblk}
\usepackage{here}
\usepackage{subfig}
\usepackage[pdftex]{graphicx}
\usepackage{color}
\usepackage{braket}
\usepackage{amssymb, amsmath}
\usepackage{feynmf}
\usepackage{bm}
\usepackage{url}
\usepackage{slashed}
\usepackage{caption}
\usepackage{tabularx}

\usepackage{jcappub}
\usepackage[compat=1.1.0]{tikz-feynhand}
\usepackage{hyperref}
\usepackage[samesize]{cancel}
\usepackage{fancyhdr} 

\newcommand{\fng}{f_{\rm NG}}

\newcommand{\beq}{\begin{equation}}
\newcommand{\eeq}{\end{equation}}

\newcommand{\lmk}{\left(}
\newcommand{\rmk}{\right)}

\begin{document}

%%
%% constraints on?
%% 

\title{\boldmath Revised bounds on local cosmic strings from NANOGrav observations}
\author[a,b,c]{Jun'ya Kume}
\author[d,e]{and Mark Hindmarsh}
\affiliation[a]{Dipartimento di Fisica e Astronomia ``G. Galilei'', Universit\`a degli Studi di Padova, via Marzolo 8, I-35131 Padova, Italy}
\affiliation[b]{INFN, Sezione di Padova, via Marzolo 8, I-35131 Padova, Italy}
\affiliation[c]{Research Center for the Early Universe (RESCEU), Graduate School of Science, The University of Tokyo, Hongo 7-3-1
Bunkyo-ku, Tokyo 113-0033, Japan}
\affiliation[d]{Department of Physics and Helsinki Institute of Physics, PL 64, FI-00014 University of Helsinki,
Finland}
\affiliation[e]{Department of Physics and Astronomy, University of Sussex, Falmer, Brighton BN1 9QH, U.K}

% e-mail addresses: one for each author, in the same order as the authors
\emailAdd{junya.kume@unipd.it}
\emailAdd{mark.hindmarsh@helsinki.fi}

\subheader{{\rm RESCEU-4/24, HIP-2024-11/TH}}

\abstract{
%Cosmic string networks, which may be formed in the early universe, are potentially a source of a stochastic gravitational wave background (SGWB) signalling new physics in the very early Universe.
In a recent paper, the NANOGrav collaboration studied new physics explanations of the observed pulsar timing residuals consistent with a stochastic gravitational wave background (SGWB)~\cite{NANOGrav:2023hvm}, including cosmic strings in the Nambu-Goto (NG) approximation. Analysing one of current models for the loop distribution, it was found that the cosmic string model is disfavored compared to other sources, for example, super massive black hole binaries (SMBHBs). When both SMBHB and cosmic string models are included in the analysis, an upper bound on a string tension $G\mu \lesssim 10^{-10}$ was derived. However, the analysis did not accommodate results from cosmic string simulations in an underlying field theory, which indicate that at most a small fraction of string loops survive long enough to emit GW. Following and extending our previous study~\cite{Hindmarsh:2022awe}, we suppose that a fraction $\fng$ of string loops follow NG dynamics and emit only GWs, and study the three different models of the loop distribution discussed in the LIGO-Virgo-KAGRA (LVK) collaboration analyses.
We re-analyse the NANOGrav 15yrs data with our signal models by using the NANOGrav \texttt{ENTERPRISE} analysis code via the wrapper \texttt{PTArcade}. We find that loop distributions similar to LVK Model B and C yield higher Bayes factor than Model A analysed in the NANOGrav paper, as they can more easily accommodate a blue-tilted spectrum of the observed amplitude. Furthermore, because of the degeneracy of $G\mu$ and $\fng$ in determining the signal amplitude, our posterior distribution extends to higher values of $G\mu$, and in some cases the uppermost value of credible intervals is close to the Cosmic Microwave Background limit $G\mu \lesssim 10^{-7}$. 
Hence, in addition to the pulsar timing array data, further information about the fraction of long-lived loops in a cosmic string network is required to constrain the string tension.
}
\maketitle
\flushbottom

\section{Introduction}
Cosmic strings are linear concentrations of energy across cosmological scales that could have formed during early Universe phase transitions in various high-energy physical scenarios~\cite{Vilenkin:278400,Hindmarsh:1994re,Hindmarsh:2011qj,Copeland:2011dx}. In the conventional description of cosmic strings, they are approximated as infinitely thin line-like objects. The strings are then expected to evolve according to the Nambu--Goto (NG) equation~\cite{Forster:1974ga,Arodz:1995dg,Anderson:1997ip} and the re-connection rule when they cross each other~\cite{Shellard:1987bv,Matzner:1988qqj,Achucarro:2006es}. 
Within this picture, the loops of cosmic string slowly decay by radiating gravitational waves (GWs) and an observable stochastic gravitational wave background (SGWB) can be generated~\cite{Auclair:2019wcv}.
Since the loop distribution is approximately scale-invariant, the cosmic string SGWB spectrum is predicted to cover a very wide frequency range. The SGWB from NG string loops has therefore been one of the primary targets of the pulsar timing arrays (PTAs)~\cite{Lentati:2015qwp,Shannon:2015ect, Blanco-Pillado:2017rnf, Ringeval:2017eww, NANOGRAV:2018hou} and laser interferometric GW detectors~\cite{Auclair:2019wcv,LIGOScientific:2021nrg}.

Recently, PTA collaborations (NANOGrav, EPTA/InPTA, PPTA and CPTA) have reported evidence for an isotropic SWGB as the expected Hellings-Downs (HD) angular correlation between pulsars’ line of sight were measured at $3-4 \sigma$ confidence level~\cite{NANOGrav:2023gor, EPTA:2023fyk, Reardon:2023gzh, Xu:2023wog}.
With high significance of the common spectrum and the strong evidence of HD correlation, it is highly motivated to investigate whether current observations can be explained by SGWB from the population of cosmic string loops~\cite{NANOGrav:2023hvm, EPTA:2023xxk, 
King:2023cgv,
Kitajima:2023vre, Ellis:2023tsl,Wang:2023len,Lazarides:2023ksx, Eichhorn:2023gat, Buchmuller:2023aus, Yamada:2023thl, Fu:2023mdu, Lazarides:2023rqf, Ahmed:2023rky,
Afzal:2023cyp, Ahmed:2023pjl, Afzal:2023kqs, King:2023wkm, Marfatia:2023fvh} (see also Refs.~\cite{Benabou:2023ghl, Baeza-Ballesteros:2023say} for global strings.). Indeed, NANOGrav and EPTA performed Bayesian inference for the models of NG loop distribution~\cite{Lorenz:2010sm,Blanco-Pillado:2013qja}, whose SGWB spectrum is controlled by the dimensionless string tension $G\mu$.
Because the NG models have difficulty reproducing the blue-tilted spectrum favored by the data with consistent amplitude, upper bounds on the string tension are derived when combining the NG models and the ``more favored'' super massive black hole binary (SMBHB) signal~\cite{NANOGrav:2023hvm, EPTA:2023xxk}. 

However, NG models of cosmic string loop evolution lack support from field theoretic simulations of loops in cosmic string networks~\cite{Vincent:1997cx,Moore:2001px,Bevis:2006mj,Bevis:2010gj,Daverio:2015nva,Correia:2019bdl,Correia:2020gkj,Correia:2020yqg,Hindmarsh:2021mnl}. 
Investigations targeting string loops produced from an evolving network show that they decay as fast as causally allowed via the radiation of classical scalar and gauge fields~\cite{Hindmarsh:2008dw,Hindmarsh:2021mnl}, even for strings whose length is O($10^3$) larger than their width where the NG approximation would be expected to hold. 
While nearly all field theory simulations of cosmic strings have been carried out in the Abelian Higgs (AH) model, which consists of a U(1) gauge field and associated scalar field, simulations of strings in an SU(2) theory show the same behaviour \cite{Hindmarsh:2018zch}.
This would imply that the primary constraint on cosmic strings is provided not by SGWB observations, but by the cosmic microwave background~\cite{Planck:2015fie, Lizarraga:2016onn} and, in the case of strings with a decay channel into Standard Model particles, by the light element abundances and the Diffuse Gamma-Ray background~\cite{SantanaMota:2014xpw}.
If dark matter is included in the decay products, it is further constrained by the dark matter relic density~\cite{Hindmarsh:2013jha}.

On the other hand, field configurations whose evolution is well approximated by NG dynamics can be created by carefully chosen initial conditions~\cite{Matsunami:2019fss,Hindmarsh:2021mnl}.
Such NG-like strings, if large enough, would decay primarily through the radiation of GWs and the observable SGWB might be produced~\cite{Hindmarsh:2022awe} while the most loops decay into the classical radiation. 
That said, no such long-lived loops are observed in simulations with the random initial conditions modelling a phase transition producing the string network.
In order to account for this theoretical uncertainty, a parameter $\fng$ quantifying the fraction of loops following NG dynamics was introduced in Refs.~\cite{Hindmarsh:2021mnl,Hindmarsh:2022awe}. 
Non-observation of such loops randomly generated initial conditions provides an upper bound as $\fng \lesssim 0.1$. 

In this work, we conduct the Bayesian inference analyses for cosmic strings using the latest NANOGrav 15yr datasets, including the parameter $\fng$ as well as the dimensionless string tension $G\mu$, in line with our previous study using their 12.5yr observational result~\cite{Hindmarsh:2022awe}. All the analyses were performed with a wrapper \texttt{PTArcade}~\cite{Mitridate_2023,Mitridate:2023oar} that allows us to easily handle \texttt{ENTERPRISE}~\cite{enterprise,enterprise-ext} where the data analysis method used in the NANOGrav collaboration is implemented.
Since both parameters control the amplitude of the SGWB, degeneracy arises and the posterior distributions can be extended to much higher $G\mu$ than in the pure NG loop scenario.  

As no NG-like loops have been observed in the AH model, the distribution of NG-like loops is completely unknown. To quantify SGWB from those loops under this circumstance, we use established models for pure NG loop distribution as an approximation. 
In particular, we use the three models studied by the LIGO-Virgo-KAGRA (LVK) collaboration , only one of which was used in Ref.~\cite{Hindmarsh:2022awe} and the NANOGrav analysis \cite{NANOGrav:2023hvm}. 
Our paper therefore extends the NANOGrav analysis to include a much wider range of models of cosmic string evolution, which, in our view, accounts much better for the theoretical uncertainties. 

The rest of the paper is organized as follows. In Sec.~\ref{sec:SGWB_from_AH}, we introduce NG loop distribution models that we refer to and discuss our characterization of the possible SGWB from NG-like loops in the AH model based on Refs.~\cite{Hindmarsh:2021mnl,Hindmarsh:2022awe}. 
Then, after presenting the summary of analyses performed with \texttt{PTArcade}, we present the results of our Bayesian analysis for each models in Sec.~\ref{sec:data_analysis}. Sec.~\ref{sec:discussion} is devoted to the discussion.

\section{Modelling the SGWB from string loops}
\label{sec:SGWB_from_AH}

\subsection{SGWB from field theory string loops}
As mentioned in the introduction, long-lived oscillating NG-like loops, required for a significant GW signal, have not yet been observed in large-scale field theory simulations~\cite{Hindmarsh:2021mnl}.  Therefore their length distribution, needed for calculations of the GW power spectrum, is completely unknown.
In Ref.~\cite{Hindmarsh:2021mnl}, (a part of) this uncertainty in their distribution was parameterised by allowing a fraction $f_{\rm NG}$ of loops to survive to radiate only gravitationally.
Subsequently in Ref.~\cite{Hindmarsh:2022awe}, it was assumed that %the NG-like loops\JKc{I think this must be all the AH loops.}
all the AH loops would have the same length distribution as in an NG network ${\sf n}(l,t)$, and hence that the distribution of NG-like loops would be $\fng {\sf n}(l,t)$.
Then the SGWB from NG-like distribution in the AH string network is quantified as 
\begin{align}
    \Omega^{\rm (AH)}_{\rm gw} = \fng \Omega^{\rm (NG)}_{\rm gw}.~\label{eq:SGWB_AH}
\end{align}
We follow this quantification and discuss possible models of NG loop distributions in the rest of the section.

We emphasise that use of the models of NG loop distribution is only a first approximation since there is no information from the AH simulations what ${\sf n}(l,t)$ should be used.
In fitting the NANOGrav data with Eq.~\eqref{eq:SGWB_AH}, the interpretation of $\fng$ as parameterising the fraction of NG-like loops holds only if the NG models are close to the true distribution of any %long-lived 
loops in a field theory network. This closeness is our underlying assumption in the following discussion.
In practice, the normalisation of ${\sf n}(l,t)$ is highly affected by the lengths of the long-lived loops at their production~\cite{Sousa:2020sxs} and also by the efficiency of loop production~\cite{Gouttenoire:2019kij}.
Therefore, even if the functional form of ${\sf n}(l,t)$ is correct, such an ambiguity in the normalisation might be still encapsulated in $\fng$.
Nevertheless, assuming that the loop size at the production well agrees with the BOS model, the authors of Ref.~\cite{Gouttenoire:2019kij} quantify the suppression of the loop production efficiency in the AH string network and the resultant overall amplitude of SGWB due to the classical radiation, which was only by a factor $\sim$ 2.
With the (optimistic) expectation that the correction in Eq.~\eqref{eq:SGWB_AH} is O(1), the development of improved models of loop distribution is left to future work.
In any case, the difference in the following results for different reference NG loop models should be understood as the underlying uncertainty in the true loop distribution.

\subsection{SGWB from NG string loops}
If the strings obey Nambu-Goto (NG) dynamics, the SGWB from the cosmic string network is dominantly sourced by the oscillation of the sub-horizon loops. 
Therefore, the number density ${\sf n}(l,t)$ of non-self-intersecting, sub-horizon cosmic string loops of invariant length $l$ at cosmic time $t$ is a necessary ingredient in evaluating the SGWB spectrum.
The present day spectrum of the SGWB can be calculated from
\begin{align}
\Omega_{\rm gw}^{\rm (NG)}(f) \equiv \frac{1}{\rho_c}\frac{d\ln\rho^{\rm (NG)}_{\rm gw}}{d\ln f} = \frac{8\pi f G^2\mu^2}{3H_0^2}\sum_{n = 1}^{\infty}C_n(f)P_n,    
\end{align}
where 
\begin{align}
    C_n(f)=\frac{2n}{f^2}\int_0^{\infty}\frac{dz}{H(z)(1+z)^6}{\sf n}\left(\frac{2n}{(1+z)f},t(z)\right)
\end{align}
with $H(z)$ and $t(z)$ being the Hubble parameter and cosmic time at redshift $z$.
The function $C_n(f)$ gives the number density of loops emitting GWs observed at frequency $f$ in the $n$-th harmonic, while the average loop power spectrum $P_n$ represents the average GW power emitted by the $n$-th harmonic of a loop. 
We note that if there is a constant factor uncertainty common to all $P_n$, then it would be absorbed by $\fng$ in~\eqref{eq:SGWB_AH} and affect the interpretation of the analyses. For simplicity, we assume that $\fng$ is solely contributed from the uncertainty in the NG-like loop distribution in the following section.

The constants $P_n$ depend on the average shape of the loops, and in particular the number of cusps per oscillation and the number of kinks travelling around the loop, which dictate the high frequency behaviour. Their sum (or the so-called decay constant of strings) $\Gamma$ can be decomposed into three contributions as
\begin{align}
    \Gamma &\equiv \sum_{n = 1}^{\infty} P_n = \sum_{n = 1}^{\infty} \lmk P_{n, {\rm c}} + P_{n, {\rm k}} + P_{n, {\rm kk}}\rmk \\
     &= N_{\rm c}\frac{3\pi^2g_{\rm 1,c}^2}{g_2^{1/3}} + N_{\rm k}\frac{3\pi^2g_{\rm 1,k}^2}{g_2^{-1/3}} + N_{\rm kk}2\pi^2g_{\rm 1,kk}^2,\label{eq:gamma_decompose}
\end{align}
where the subscripts ${\rm c, k, kk}$ represents the contributions from cusps, kinks and kink-kink collisions, and $P_n$ scales as $n^{-4/3}, n^{-5/3},$ and $n^{-2}$ respectively. 
In the second line, $N_{\rm c}$ is the average number of cusps per oscillation, $N_{\rm k}$ the average number of kinks per loop, and $N_{\rm kk}$ is the number of kink-kink collisions per oscillation. In the limit of large number of kinks, $N_{\rm kk} \to N_{\rm k}^2/4$, and we adopt the limiting value in our models. The other numerical factors are given as $g_{\rm 1,c} = 0.85$, $g_{\rm 1,k} = 0.29$, $g_{\rm 1,kk} = 0.10$ and $g_2 = \sqrt{3}/4$. 
Note that in the numerical simulations of individual loops, $\Gamma \simeq 50$ is found~\cite{Blanco-Pillado:2017oxo}.
As a benchmark, we consider the following three models of NG loops in our analysis. 

\subsubsection{BOS model (LVK-Model A)}
In this model, based on NG simulations of string networks in the radiation and matter dominated eras in Ref.~\cite{Blanco-Pillado:2013qja}, the number density of non-self-intersecting loops is analytically inferred from the loop production function obtained in the simulation. 
The distribution functions are different according to the cosmological era in which the loops were produced and in which they radiate in the frequency of interest, and are 
\begin{align}
{\sf n}_{\rm r,r}(l,t)&=\frac{0.18}{t^{3/2}(l + \Gamma G\mu t)^{5/2}}\Theta(0.1-l/t),\label{rr}\\
{\sf n}_{\rm r,m}(l,t)&=\frac{0.18(2H_0\sqrt{\Omega_r})^{3/2}(1+z)^3}{(l + \Gamma G\mu t)^{5/2}}\Theta(0.09t_{\rm eq}/t-\Gamma G\mu -l/t),\label{rm}\\
{\sf n}_{\rm m,m}(l,t)&=\frac{0.27-0.45(l/t)^{0.31}}{t^{2}(l + \Gamma G\mu t)^{2}}\Theta(0.18-l/t),\label{mm}
\end{align}
where $H_0$ is the Hubble constant, $\Omega_{\rm r}$ is the density parameter of the radiation, $t_{\rm eq}$ represents the cosmic time of radiation-matter equality. 
The subscript ``r,m", for example, represents the loops produced in the radiation era and emitting GWs in matter era. 
Note that Eq.~\eqref{rm} matches to Eq.~\eqref{rr} in the early radiation era.

As a power spectrum of loops $P_n$, the ``smoothed" model~\cite{Blanco-Pillado:2017oxo} is often adopted for this BOS distribution (see {\it e.g.} the analysis of Refs.~\cite{Blanco-Pillado:2021ygr,NANOGrav:2023gor}.
It is constructed from the numerical simulation taking into account the gravitational backreaction and different from the simple decomposition in Eq.~\eqref{eq:gamma_decompose}.
Following the results of their simulation, we set $\Gamma = 50$ and extract the values of $P_n$ from the Figure~3 and~4 of Ref.~\cite{Blanco-Pillado:2017oxo}.
As another example in this class of power spectra, we also consider the kink-dominated case $(N_c, N_k) = (1, 100)$ with Eq.~\eqref{eq:gamma_decompose}, which yields $\Gamma = 710$.
Notice that the value of $\Gamma$, which affects the amplitude of SGWB, is totally different from those in Ref.~\cite{Blanco-Pillado:2017oxo}. Given the uncertainty on the initial number of kinks at the loop production, however, this case was analysed in Refs.~\cite{LIGOScientific:2021nrg, EPTA:2023hof, EPTA:2023fyk}. Therefore, these choices allow for a direct comparison of our results below with those of previous analyses.

In the NANOGrav collaboration paper~\cite{NANOGrav:2023hvm}, the BOS model with the smoothed power spectrum model was dubbed as ``STABLE-N'' model and subjected to their analysis for the new physics interpretation. It is, however, disfavored compared to the other new physics models and the SMBHB signal.
This is because the NANOGrav 15yr data favors the blue-tilted SGWB spectrum. For the BOS model, the spectral tilt becomes blue when the string tension becomes smaller, which then yields magnitude of SGWB too small to explain the observed power excess.

\subsubsection{LRS model (LVK-Model B)}
The second model is based a different set of NG simulations~\cite{Lorenz:2010sm}. In contrast with the BOS model, the distribution of large non-self-interacting loops ($l/t > \gamma_d \equiv \Gamma G\mu$) is directly extracted from the simulation. Its distribution is extended down to the smaller size by solving the Boltzmann equation with a theoretically derived loop production function~\cite{Polchinski:2006ee}. Note that this loop production function introduces the new scale $\gamma_c (< \gamma_d)$ corresponding to the scale of gravitational backreaction. The analytical approximate expression of the LRS loop distribution function depends on the regimes of loop length as

\begin{equation}
t^4{\sf n}_{\rm r,r}(l,t) =
\left\{ \,
\begin{aligned}
&\frac{0.08}{(l/t + \Gamma G\mu)^{3 - 2\chi_r}} \quad ({\rm for\ }l/t > \Gamma G\mu ),\\
&\frac{0.08(1/2 - 2\chi_r)}{(2 - 2\chi_r)\Gamma G\mu (l/t)^{2 - 2\chi_r}}  \quad ({\rm for\ } \gamma_c < l/t < \Gamma G\mu ),\\
&\frac{0.08(1/2 - 2\chi_r)}{(2 - 2\chi_r)\Gamma G\mu \gamma_c^{2 - 2\chi_r}}  \quad ({\rm for\ } l/t < \gamma_c),
\end{aligned}
\right.\label{LRS_rr}
\end{equation}

\begin{equation}
t^4{\sf n}_{\rm m,m}(l,t) =
\left\{ \,
\begin{aligned}
&\frac{0.015}{(l/t + \Gamma G\mu)^{3 - 2\chi_m}} \quad ({\rm for\ }l/t > \Gamma G\mu ),\\
&\frac{0.015(1/2 - 2\chi_m)}{(2 - 2\chi_m)\Gamma G\mu (l/t)^{2 - 2\chi_m}}  \quad ({\rm for\ } \gamma_c < l/t < \Gamma G\mu ),\\
&\frac{0.015(1/2 - 2\chi_m)}{(2 - 2\chi_m)\Gamma G\mu \gamma_c^{2 - 2\chi_m}}  \quad ({\rm for\ } l/t < \gamma_c),
\end{aligned}
\right.\label{LRS_mm}
\end{equation}

\begin{align}
{\sf n}_{\rm r,m}(l,t)&=\lmk\frac{t}{t_{\rm eq}}\rmk^4 \lmk\frac{1 + z}{1 + z_{\rm eq}}\rmk^3 t_{\rm eq}^4{\sf n}_{\rm r,r}\lmk\frac{l + \Gamma G\mu(t - t_{\rm eq})}{t_{\rm eq}}\rmk,\label{LRS_rm}
\end{align}
where $(\chi_r, \chi_m) = (0.2, 0.295)$ and $\gamma_c \simeq 20(G\mu)^{1 + 2\chi{r/m}}$.
The most important difference between the BOS model and the LRS model is the dominance of smaller loops in the distribution for the latter model. Therefore, the amplitude of SGWB becomes larger in the higher frequency range and the ground-based GW detectors put the strongest bound on the string tension for this model~\cite{LIGOScientific:2021nrg}.

In the context of recent PTA observation, this model was analysed (together with the BOS model) by the EPTA collaboration with their second data release~\cite{EPTA:2023xxk}. 
Although they do not perform a model selection analysis, the posterior distribution they obtained somewhat supports the NANOGrav result mentioned above. That is, when they include the SMBHB signal in their analysis, they observe the posterior distribution for $G\mu$ extends downward to the limit of the prior (see the right panel of Fig.~\ref{fig:NG_BOS} in App.~\ref{app:exact_BOS} or Fig.~15 of Ref.~\cite{EPTA:2023xxk}). 
This indicates that with the addition of the SMBHB signal, which is more preferred by the data, the magnitude of string signal (or $G\mu$) could only be bounded above.

Following the analyses performed in Refs.~\cite{EPTA:2023hof, EPTA:2023xxk}, here we consider two different scenarios for the loop power spectrum.
The first one is $(N_c, N_k) = (2, 0)$, which yields $\Gamma = 57$, a value close to that observed in the simulations.
Another one is again $(N_c, N_k) = (1, 100)$ allowing for the large number of kinks.
Here again, our choice is intended as a comparison with the results of those previous studies. If the smoothed model is adopted instead of the first model, qualitatively similar results are expected due to the dominance of cusp and the close value of $\Gamma$ (see Ref.~\cite{Blanco-Pillado:2021ygr} for comparison of different models of loop power spectrum).

\subsubsection{LVK-Model C}
The last model interpolates between the two models described above~\cite{Auclair:2019zoz, LIGOScientific:2021nrg}.
The analytic expression of loop distribution is given as 
\begin{align}
t^4{\sf n}_{\rm r,r}(l,t) = \frac{c_r}{1/2 -2\chi_r}
\times\left\{ \,
\begin{aligned}
&(l/t + \Gamma G\mu)^{2\chi_r - 3} - \frac{\gamma_{\infty}^{2\chi_r - 1/2}}{(l/t + \Gamma G\mu)^{5/2}} \quad ({\rm for\ }l/t > \Gamma G\mu ),\\
&\frac{(l/t)^{2\chi_r - 2}}{(2 - 2\chi_r)\Gamma G\mu} - \frac{\gamma_{\infty}^{2\chi_r - 1/2}}{(l/t + \Gamma G\mu)^{5/2}} \quad ({\rm for\ } \gamma_c < l/t < \Gamma G\mu ),\\
&\frac{\gamma_c^{2\chi_r - 2}}{(2 - 2\chi_r)\Gamma G\mu} - \frac{\gamma_{\infty}^{2\chi_r - 1/2}}{(l/t + \Gamma G\mu)^{5/2}}  \quad ({\rm for\ } l/t < \gamma_c),
\end{aligned}
\right.\label{Model_C_rr}
\end{align}

\begin{equation}
t^4{\sf n}_{\rm m,m}(l,t) =\frac{c_m}{1/2 -2\chi_m}
\times\left\{ \,
\begin{aligned}
&(l/t + \Gamma G\mu)^{2\chi_m - 3} - \frac{\gamma_{\infty}^{2\chi_m - 1/2}}{(l/t + \Gamma G\mu)^{5/2}} \quad ({\rm for\ }l/t > \Gamma G\mu ),\\
&\frac{(l/t)^{2\chi_m - 2}}{(2 - 2\chi_m)\Gamma G\mu} - \frac{\gamma_{\infty}^{2\chi_m - 1/2}}{(l/t + \Gamma G\mu)^{5/2}} \quad ({\rm for\ } \gamma_c < l/t < \Gamma G\mu ),\\
&\frac{\gamma_c^{2\chi_m - 2}}{(2 - 2\chi_m)\Gamma G\mu} - \frac{\gamma_{\infty}^{2\chi_m - 1/2}}{(l/t + \Gamma G\mu)^{5/2}}  \quad ({\rm for\ } l/t < \gamma_c),
\end{aligned}
\right.\label{Model_C_mm}
\end{equation}
where $\gamma_{\infty} = 0.1$ represents the size of the largest loops in the scaling unit.
The distribution of loops produced in the radiation era and emitting GWs in matter era can be evaluated in the same way as Eq.~\eqref{LRS_rm}.
Here $c_{r/m}$ and $\chi_{r/m}$ are the model parameters.
In Ref.~\cite{LIGOScientific:2021nrg}, two sets of benchmark values are used for C-1 and C-2 scenario respectively. The former connects the BOS loop distribution in the radiation-dominated era and the LRS loop distribution in the matter dominated era, while the latter does the opposite.

In the frequency window of PTA and in the parameter regime of our interest, SGWB spectrum is dominantly determined by ${\sf n}_{\rm m,m}(l,t)$ and partly by ${\sf n}_{\rm r,m}(l,t)$.
Compared to the BOS model, the LRS model predicts a blue-tilted spectrum in this frequency band for a wider range of $G\mu$ values. 
Therefore, with our modeling of the SGWB from AH strings discussed below, it is to be expected that LRS model will be able to better explain the data.
For this practical reason, we adopt the former scenario (C-1) with $(c_r, \chi_r, c_m, \chi_m) = (0.15, 0.45, 0.019, 0.295)$ in our study. We consider the same parameters $(N_{\rm c}, N_{\rm k})$ for the loop power spectrum as in the LRS model, and take the large-$N_{\rm k}$ value for $N_{\rm kk}$.

.

\section{Data analysis of NANOGrav 15yrs observation}\label{sec:data_analysis}
\subsection{Summary of the analyses}
The Bayesian inference analyses performed by NANOGrav collaboration were implemented into \texttt{ENTERPRISE}~\cite{enterprise,enterprise-ext} via a wrapper called \texttt{PTArcade}~\cite{Mitridate_2023,Mitridate:2023oar}. 
Here we utilize this wrapper to conduct the Bayesian inference for our cosmic string model in the same way as done in the NANOGrav collaboration.
That is, we performed Markov chain Monte Carlo sampling to derive marginalized posteriors for our AH model parameters $(G\mu, \fng)$ with the timing residual data $\Vec{\delta t}$, using $30$ frequency bins for inferring pulsar intrinsic red noise and $14$ bins for common red noise.
For each analysis, $\sim 2\times10^6$ draws are generated and thinned by a factor of 10 in order to reduce the auto-correlation, and the first 25\% of each chain is burned-in.
Depending on the model, we increase the number of chains to ensure the convergence of the result.
In deriving a credible interval of the string parameters, we made use of the automated estimation of the highest probability density interval. This is, however, only applicable when the marginalized distribution is monotonic or (nearly) mono-modal.
If it is not applicable, for example, when a peak exists on top of the plateau with fluctuations or when a bi-modal structure appears, we set the interval(s) by ourselves to collect as many points as possible with a probability density above a certain level.
As summarized in Appendix.~\ref{app:exact_BOS}, we performed a test analysis with the pure BOS-model (by fixing $\fng = 1$), finding good agreement with the result of the STABLE-N model in Ref.~\cite{NANOGrav:2023hvm}.

% At the same time, 
We also perform the model selection analysis implemented in \texttt{PTArcade}, which evaluates the Bayes factor defined as 
\begin{equation}
    \mathcal{B}_{10} = \frac{p(\Vec{\delta t}|\mathcal{H}_1)}{p(\Vec{\delta t}|\mathcal{H}_0)}.
\end{equation}
Here $\mathcal{H}_1$ denotes the model under consideration, and  $\mathcal{H}_0$ the reference model, which contains the signal from SMBHB only. The SMBHB power spectrum model is expressed as~\cite{Mitridate:2023oar}
\begin{equation}
    h^2 \Omega_{\rm gw}^{\rm (SMBHB)} = \frac{2\pi^2 A_{\rm BHB}^2}{3H_0^2}\lmk \frac{f}{f_{\rm yr}}\rmk^{5 - \gamma_{\rm BHB}}f_{\rm yr}^2,\label{eq:SMBH_signal}
\end{equation}
where $H_0$ is the Hubble constant at present and $f_{\rm yr} = ({\rm year})^{-1}$. 
For binaries in circular Keplerian orbits, $\gamma_{\rm BHB} = 13/3$.
The evaluation of Bayes factor is based on product space methods~\cite{2cbc2763-3bd1-3200-951e-b1bea0d1381f, doi:10.1198/10618600152627924, Hee:2015eba} and its statistical uncertainty is derived by the bootstrapping method~\cite{Efron:1986hys}.

\begin{figure}[htbp] 
\centering
\includegraphics[clip,width=0.8\columnwidth]{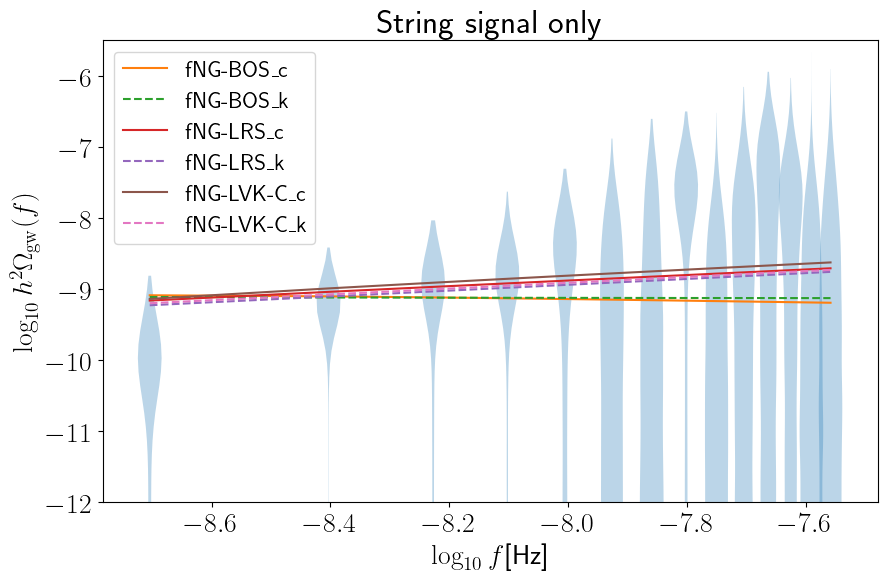}
  \caption{SGWB spectrum in our models of AH string loops with the parameters evaluated by Bayesian inference. For each model, \_k stands for the kink dominated loop spectrum.
  For reference, we also show the violin plot derived from free spectrum fit of the NANOGrav data.}
    \label{fig:Bayes_est}
\end{figure}

For each model of the NG-like loop distribution, we consider two sets of different loop power spectrum mentioned above, and perform Bayesian inference for two cases where only the SGWB signal from AH strings exists and where it can be mixed with the SMBHB signal. Prior distributions of physical parameters for AH string loop distribution and the SMBHB signal are summarized in Table~\ref{tab:prior}.
In Appendix.~\ref{app:prior}, we discuss how our result could be affected by changing the prior for the SMBHB (or power-law) signal.
Outputs of analysis, namely, Bayesian Estimators, Maximum Posteriors, Credible Intervals for the parameters and Bayes factors over SMBHB interpretation are summarized for each model in Tables~\ref{tab:post_BOS}--~\ref{tab:post_Model_C}.
In Fig.~\ref{fig:Bayes_est}, we plot the signals in our model for the Bayes estimator of parameters.
Note that the prior for string tension in our analysis is the same as that used in Ref.~\cite{NANOGrav:2023hvm}, which is restricted to the parameter region where models produce an observable signal in the PTA band.
Here the prior of $\fng$ was set with the same spirit.

As discussed in Refs.~\cite{NANOGrav:2023hvm, EPTA:2023xxk}, however, the information carried by the ``signal'' is not particularly constraining at the moment and the result should crucially depend on the priors. 
Therefore, especially under the significant uncertainty in the underlying AH loop distribution, too much emphasis should not be placed on the exact number of Bayes factors reported in this study.
Nevertheless, it can be considered as a good indicator for comparing which of the models we analysed is relatively better at explaining the data.

\begin{table}[H]
\caption{Prior distribution for the AH string parameters and the SMBHB signal used in this study.}\label{tab:prior}
\begin{tabular}{ccc}
 \\[-1ex]
 \hline
 Parameter & Description & Prior \\
 \hline\hline\\[-1ex]
 \multicolumn{3}{c}{\textbf{AH strings}}\\[1.5ex] 
 $G\mu$ & string tension & log-uniform [$-14$, $-6$] \\[1ex] 
 $\fng$ & fraction of NG-like loops & log-uniform [$-6$, $0$]\\[1ex] 
 \hline\\[-1ex]
 \multicolumn{3}{c}{\textbf{SMBHB signal}}\\[1.5ex] 
 ($\log_{10}A_{\rm BHB}, \gamma_{\rm BHB}$) & \begin{tabular}{c}
 Amplitude and tilt of\\ the SGWB spectrum
 \end{tabular}
 & 
 \begin{tabular}{c}
 2D normal distribution with\\
 mean $\bm{\mu}_{\rm BHB} = (-15.6, 4.7)$ and covariance \\
 $\bm{\sigma}_{\rm BHB} = \begin{pmatrix}0.28 & -2.6\times10^{-3} \\ -2.6\times10^{-3} & 0.12 \end{pmatrix}$
\end{tabular}
 \\[-1.5ex] \\
 \hline
\end{tabular}
\end{table}

\subsection{Results for each model}
In the following, we describe the results of our Bayesian inference analyses for each model. For the names of models, we use the following notation:
\begin{align}
  \texttt{Model\ Name} = \texttt{fNG-}\langle \texttt{NG\ model}\rangle\texttt{\_}\langle \texttt{shape}\rangle \lmk + \texttt{SMBHB}\rmk, \notag
\end{align}
where $\langle \texttt{NG\ model}\rangle =$ \texttt{BOS, LRS, LVK-C} and $\langle \texttt{shape}\rangle =$ \texttt{c}, \texttt{k} (cusp or kink-dominated respectively\footnote{Since the smoothed loop power spectrum for the BOS model assumes that only cusps exist, we shall refer to this case as BOS\_c.}). The last part $\lmk + \texttt{SMBHB}\rmk$ indicates whether the SMBHB signal is superimposed or not. 

\subsubsection{$\fng$-BOS models}
\begin{figure}[htbp] 
\centering
\includegraphics[clip,width=0.5\columnwidth]{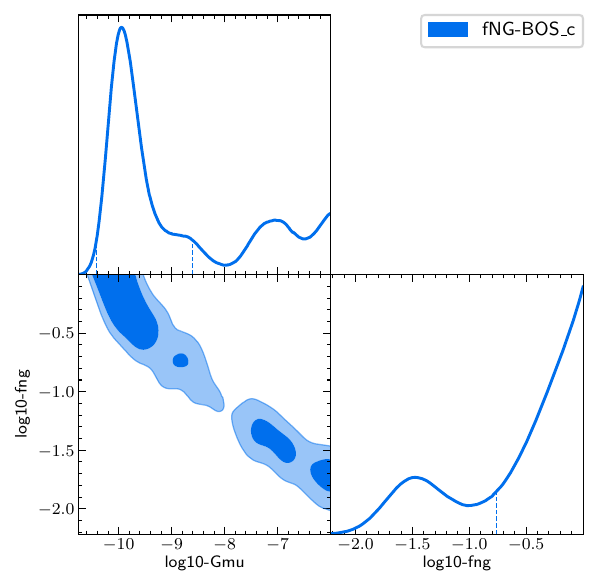}~
\includegraphics[clip,width=0.5\columnwidth]{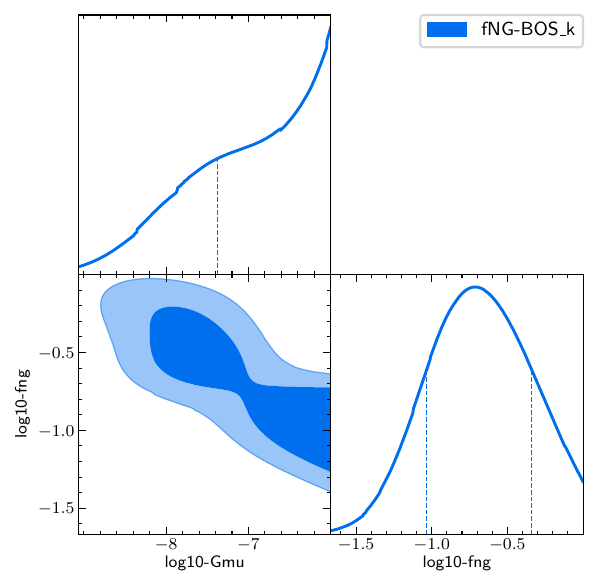}
  \caption{Posterior distribution of the string parameters for BOS-like models without SMBHB signal. The left panel is for the smoothed loop distribution and the right one is for kink dominated model.}
    \label{fig:BOS_fng}
\end{figure}

Let us start from the case where only the string signal is assumed.
In Fig.~\ref{fig:BOS_fng}, we plot the posterior distribution of the string parameters for the BOS-like distribution. The left panel is for the smoothed loop power spectrum.
As one can see, there are two favored branches $(G\mu, \fng) \simeq (10^{-10},1)$ and $(G\mu, \fng) \simeq (10^{-6}, 10^{-1.5})$ . Since the lower $G\mu$ branch seems to be more favored, we estimate the 68\% credible intervals considering only the contribution of this branch.
This branch, however, cannot be allowed when the upper bound on $\fng \lesssim 0.1$ derived from the simulation~\cite{Hindmarsh:2021mnl} is naively applied. On the other hand, the string tension is gravitationally constrained by CMB as $G\mu \lesssim 10^{-7}$, and therefore both favored branch seems to be incompatible with existing limits.
The right panel is for the kink dominated spectrum. Because the amplitude of SGWB is highly suppressed for the BOS model in this case, the posterior distribution is concentrated in a very narrow range.
Also in this case, relatively higher values of $G\mu$ and $\fng$ are required to explain the data and might be contradict with those bounds.

\begin{figure}[htbp] 
\centering
\includegraphics[clip,width=0.5\columnwidth]{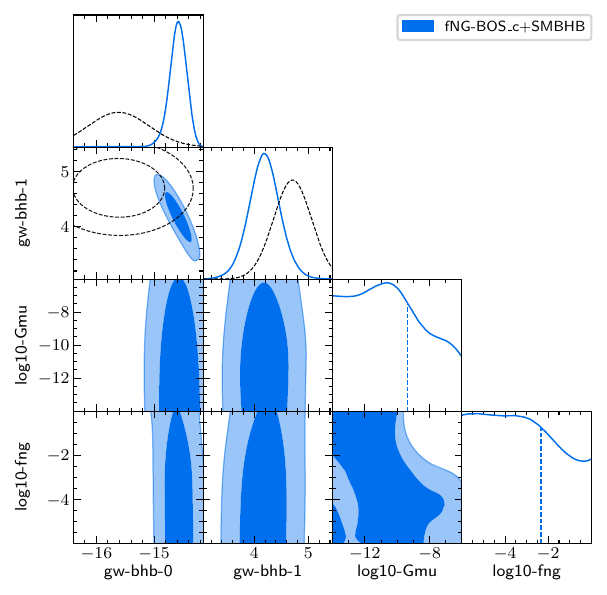}~
\includegraphics[clip,width=0.5\columnwidth]{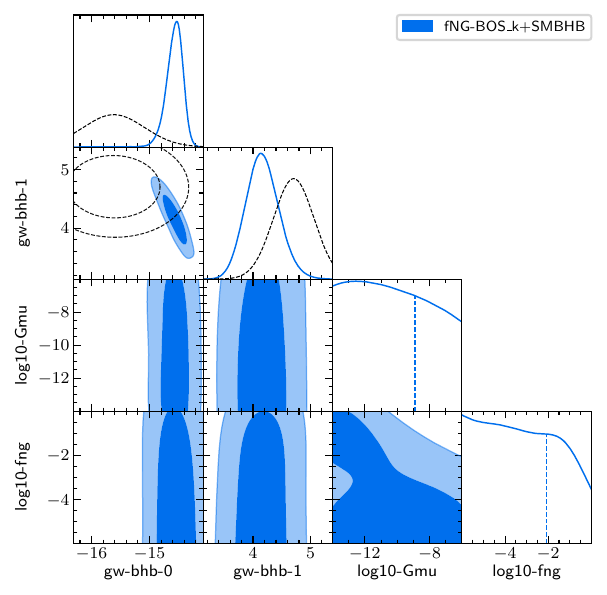}
  \caption{Posterior distribution of the string parameters for BOS-like models with SMBHB signal. The left and right panels are for the same loop spectrum as Fig.~\ref{fig:BOS_fng}.}
    \label{fig:BOS_fng_SMBH}
\end{figure}

We also analyse the case where the indicated SGWB signal is the superposition of the signal from SMBHB and the AH strings. In Fig.~\ref{fig:BOS_fng_SMBH}, the posterior distribution of the string parameters and SMBHB is plotted. 
Hereafter in the posterior plots, {\texttt gw-bhb-0} and {\texttt gw-bhb-1} represent $A_{\rm BHB}$ and $\gamma_{\rm BHB}$ respectively.  
Again, the left panel is for the smoothed loop power spectrum and the right panel is for the kink dominated spectrum.
In both cases, it can be seen that the data favours the sub-dominance of the string signal and the credible intervals of $(G\mu, \fng)$ are bounded from above.
As expected, the upper most value of $G\mu$ in the credible interval becomes larger than the case with pure BOS model $[-11.99, -9.90]$ (see Table~4 in Ref.~\cite{NANOGrav:2023hvm}).
In other words, higher $G\mu$ (or symmetry breaking scales) are to some extent still allowed for by taking into account that not all string loops emit GWs.

Overall, the Bayes factor results in smaller values ($\mathcal{B}_{10} < 1$) indicating that in all cases the present model is less favoured than the SMBHB signal.
In particular, the string-only case results in the even smaller Bayes factor than in the case of the pure BOS model due to the parameter space extended by $\fng$.
Nevertheless, a possibly interesting implication from the superposition case is that the addition of string signal contribution makes the posterior distribution of the SMBHB signal parameters somewhat more consistent with the prior distribution derived with numerical simulations. 
Such a behavior was also observed in the analyses by NANOGrav collaboration~\cite{NANOGrav:2023hvm}.

\begin{table}[htbp]
\caption{Summary of the result for BOS-like distribution. Bayesian Estimators, Maximum Posteriors, 68\% Credible Intervals for the string parameters and Bayes factor (BF) over SMBHB interpretation are listed. Values with $^*$ are at the edge of the range assumed in the prior distribution.
}\label{tab:post_BOS}
\begin{tabular}{ccccc}
 \\[-1ex]
 \multicolumn{5}{c}{\textbf{$\fng$-BOS models}}\\[0.5ex] 
 \hline
 Parameter & Bayes Est. & Maximum Post. & 68\% Credible Interval & BF\\ [0.5ex] 
 \hline\hline\\[-1ex]
 \multicolumn{5}{c}{\textbf{String with smoothed loop model}}\\[1ex] 
 \begin{tabular}{c}
 $\log_{10}G\mu$ \\
 $\log_{10}\fng$
 \end{tabular}
 &
 \begin{tabular}{c}
 $-8.96 \pm 1.35$ \\
 $-0.64 \pm 0.57$
 \end{tabular}
 &
 \begin{tabular}{c}
 $-9.94$ \\
 $0^*$
 \end{tabular}
 &
 \begin{tabular}{c}
 $[-10.42, -8.61]$ \\
 $[-0.76, 0^*]$
 \end{tabular}
 & 0.014\\ \\[-1.5ex]
 \hline\\[-1ex]
  \multicolumn{5}{c}{\textbf{String with $(N_c, N_k) = (1, 100)$}}\\[1ex] 
 \begin{tabular}{c}
 $\log_{10}G\mu$ \\
 $\log_{10}\fng$
 \end{tabular}
 &
 \begin{tabular}{c}
 $-7.04 \pm 0.73$ \\
 $-0.71 \pm 0.29$
 \end{tabular}
 &
 \begin{tabular}{c}
 $-6^*$ \\
 $-0.71$
 \end{tabular}
 &
 \begin{tabular}{c}
 $[-7.37, -6^*]$ \\
 $[-1.03, -0.33]$
 \end{tabular}
 & 0.0043\\ \\[-1.5ex]
 \hline\\[-1ex]
 \multicolumn{5}{c}{\textbf{SMBHB + String with smoothed loop model}}\\[1ex] 
 \begin{tabular}{c}
 $\log_{10}G\mu$ \\
 $\log_{10}\fng$
 \end{tabular}
 &
 \begin{tabular}{c}
 $-10.43 \pm 2.15$ \\
 $-3.24 \pm 1.67$
 \end{tabular}
 &
 \begin{tabular}{c}
 $-10.69$ \\
 $-5.43$
 \end{tabular}
 &
 \begin{tabular}{c}
 $[-14^*, -9.36]$ \\
 $[-6^*, -2.34]$
 \end{tabular}
 & 0.82\\ \\[-1.5ex]
 \hline\\[-1ex]
  \multicolumn{5}{c}{\textbf{SMBHB + String with $(N_c, N_k) = (1, 100)$}}\\[1ex] 
 \begin{tabular}{c}
 $\log_{10}G\mu$ \\
 $\log_{10}\fng$
 \end{tabular}
 &
 \begin{tabular}{c}
 $-10.22\pm 2.25$ \\
 $-3.23 \pm 1.67$
 \end{tabular}
 &
 \begin{tabular}{c}
 $-12.5$ \\
 $-6^*$
 \end{tabular}
 &
 \begin{tabular}{c}
 $[-14^*, -8.90]$ \\
 $[-6^*, -2.08]$
 \end{tabular}
 & 0.98\\ \\[-1.5ex]
 \hline\\[-1ex]
\end{tabular}
\end{table}

\subsubsection{$\fng$-LRS models}
\begin{figure}[htbp] 
\centering
  \includegraphics[clip,width=0.5\columnwidth]{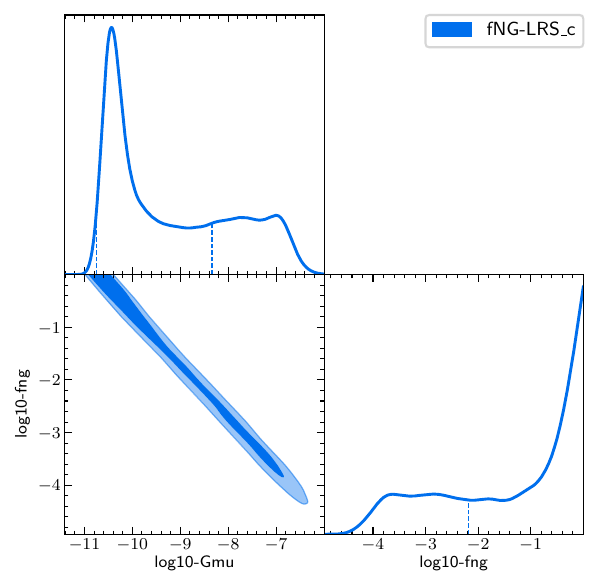}~ \includegraphics[clip,width=0.5\columnwidth]{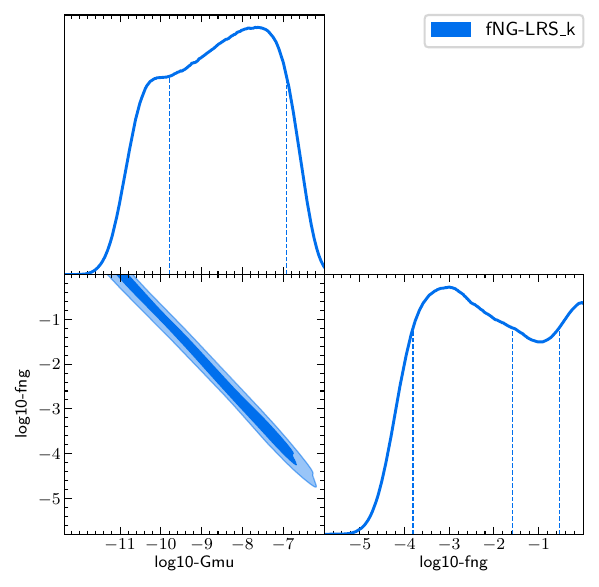}
  \caption{Posterior distribution of the string parameters for LRS-like models without SMBHB signal. For the loop power spectrum, $(N_c, N_k) = (2, 0)$ is assumed in the left panel and $(N_c, N_k) = (1, 100)$ in the right panel.}
    \label{fig:LRS_fng}
\end{figure}

Let us again begin with the case of string signal only. In Fig.~\ref{fig:LRS_fng}, the posterior distribution of string parameters are shown. 
In the present model, $(N_c, N_k) = (2, 0)$ is assumed for the loop power spectrum in the left panel while $(N_c, N_k) = (1, 100)$ in the right panel.
Compared to the results for the BOS-like loop distribution, this model predicts sharper support in the posterior distribution and yields the higher Bayes factor.
Indeed, the sharpness of the support in the pure LRS model compared to the pure BOS model was observed in Ref.~\cite{EPTA:2023xxk}, and our posterior distribution can be regarded as a stretched version of that.
The most striking difference between the different loop spectrum is that $(N_c, N_k) = (2, 0)$ case admits the concentration of samples around $(G\mu, \fng) \simeq (10^{-10},1)$. 
Nevertheless, the 2D posterior distribution is not as markedly changed by the dominance of kink as in the case of BOS, which can be understood from the lack of noticeable differences in the SGWB spectra as discussed in, for example, Refs~\cite{EPTA:2023hof,EPTA:2023xxk}.

\begin{figure}[htbp] 
\centering
\includegraphics[clip,width=0.5\columnwidth]{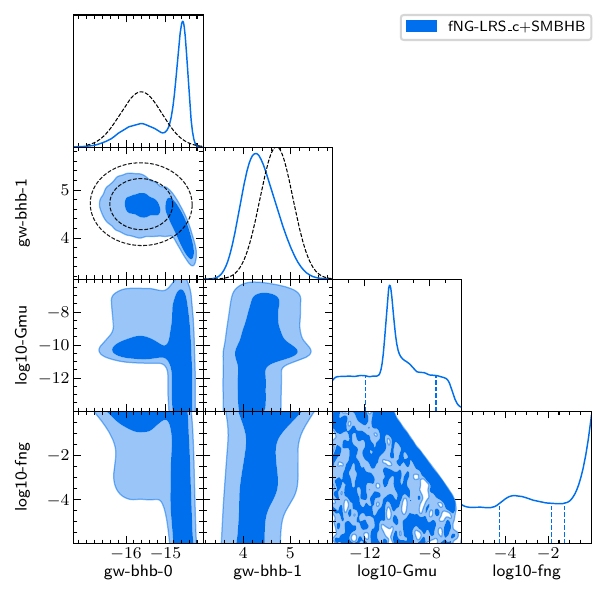}~
\includegraphics[clip,width=0.5\columnwidth]{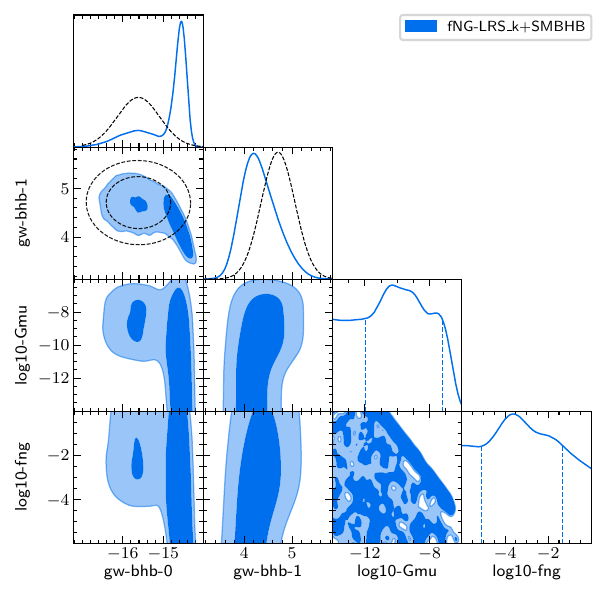}
  \caption{Posterior distribution of the string parameters for LRS-like models with SMBHB signal. The left and right panels are for the same loop power spectrum as Fig.~\ref{fig:LRS_fng}.}
    \label{fig:LRS_fng_SMBH}
\end{figure}

In Fig.~\ref{fig:LRS_fng_SMBH}, we show the posterior distribution for the cases where the superposition of string signal and SMBHB signal is assumed.
These again result in the higher Bayes factor than the same scenarios for the BOS-like distribution.
As for the string parameters, one can see the spreading of narrow support seen in the string only case and a sparse distribution to the lower left of the $(G\mu, \fng)$ plane, where the signal from the string is weaker.
The extended distribution again indicates that when one takes into account the indication of AH simulation, the data allows higher $G\mu$ (if accompanied by smaller $\fng$) than those for the pure NG loop models.
In contrast to the case with BOS-like loop distribution, one can see that certain fraction of samples remain in the range of the posterior distribution of SMBHB signal parameters. This indicates a split of the posterior into two branches, one where the SMBHB signal is dominant and the other where the string is dominant. Given that the LRS-like loop distribution is preferred by the data over the BOS-like case, it is natural that such a distribution should appear. The latter branch is also located in the vicinity of the prior as in the BOS model.
Therefore, the addition of the string signal again has the effect of bringing the posterior distribution of the SMBHB parameters closer to the prior distribution obtained from the simulation.

Now let us discuss the existing limits on the model.
For this model, one may think of the LVK constraint on string tension $G\mu$ derived at the higher frequency. As discussed in Refs.~\cite{Auclair:2019jip,Auclair:2021jud}, however, the classical radiation of the fields becomes effective for the NG-like loops when they become shorter than a critical length. This avoids the LVK constraint by imposing a high frequency cut-off on the SGWB spectrum.
By applying $\fng \lesssim 0.1$, it is required for the model to have a large $G\mu$ (and a small $\fng$) in order to explain the data (dominantly) with the AH string signal.
This makes the $\Gamma \simeq 50$ scenario (the left panel of Figs.~\ref{fig:LRS_fng} and~\ref{fig:LRS_fng_SMBH}) less-likely compared to the kink-dominated scenario, since $(G\mu, \fng) \simeq (10^{-10},1)$ is the maximum posterior.

While the constraint on $G\mu$ from SGWB become less effective when $\fng$ becomes smaller, one can instead consider the effects of particle radiation from strings.
As discussed in Refs.~\cite{SantanaMota:2014xpw, Hindmarsh:2022awe}, higher $G\mu$ might cause the significant amount of energy injection into the visible sector depending on the coupling, and can be constrained from the Big-Bang nucleosynthesis and from the diffused gamma-ray background.

 \begin{table}[htbp]
\caption{The same table as Table~\ref{tab:post_BOS} but for the LRS-like loop distribution.}\label{tab:post_LRS}
\begin{tabular}{ccccc}
 \\[-1ex]
 \multicolumn{5}{c}{\textbf{$\fng$-LRS models}}\\[0.5ex] 
 \hline
 Parameter & Bayes Est. &Maximum Post. & 68\% Credible Interval(s) & BF\\ [0.5ex] 
 \hline\hline\\[-1ex]
 \multicolumn{5}{c}{\textbf{String with $(N_c, N_k) = (2, 0)$}}\\[1ex]
 \begin{tabular}{c}
 $\log_{10}G\mu$ \\
 $\log_{10}\fng$
 \end{tabular}
 &
 \begin{tabular}{c}
 $-9.22 \pm 1.31$ \\
 $-1.48 \pm 1.31$
 \end{tabular}
 &
\begin{tabular}{c}
 $-10.47$ \\
 $0^*$
 \end{tabular}
 &
 \begin{tabular}{c}
 [$-10.75, -8.34$] \\
 $[-2.18, 0^*]$
 \end{tabular}
 & 0.56\\ \\[-1.5ex]
 \hline\\[-1ex]
 \multicolumn{5}{c}{\textbf{String with $(N_c, N_k) = (1, 100)$}}\\[1ex] 
 \begin{tabular}{c}
 $\log_{10}G\mu$ \\
 $\log_{10}\fng$
 \end{tabular}
 &
 \begin{tabular}{c}
 $-8.66 \pm 1.30$ \\
 $-2.24 \pm 1.27$
 \end{tabular}
 &
 \begin{tabular}{c}
 $-7.69$ \\
 $-3.00$
 \end{tabular}
 &
 \begin{tabular}{c}
 [$-9.78, -6.92$] \\
 $[-3.81, -1.59], [-0.54, 0^*]$
 \end{tabular}
 & 0.31\\ \\[-1.5ex]
 \hline\\[-1ex]
 \multicolumn{5}{c}{\textbf{SMBHB + String with $(N_c, N_k) = (2, 0)$}}\\[1ex]
 \begin{tabular}{c}
 $\log_{10}G\mu$ \\
 $\log_{10}\fng$
 \end{tabular}
 &
 \begin{tabular}{c}
 $-10.23 \pm 1.92$ \\
 $-2.64 \pm 1.81$
 \end{tabular}
 &
 \begin{tabular}{c}
 $-10.46$ \\
 $0^*$
 \end{tabular}
 &
 \begin{tabular}{c}
 $[-11.98, -7.60]$ \\
 $[-4.25, -1.85], [-1.24, 0^*]$
 \end{tabular}
 & 1.24\\ \\[-1.5ex]
 \hline\\[-1ex]
 
 \multicolumn{5}{c}{\textbf{SMBHB + String with $(N_c, N_k) = (1, 100)$}}\\[1ex] 
 \begin{tabular}{c}
 $\log_{10}G\mu$ \\
 $\log_{10}\fng$
 \end{tabular}
 &
 \begin{tabular}{c}
 $-10.21 \pm 2.06$ \\
 $-3.10 \pm 1.63$
 \end{tabular}
 &
 \begin{tabular}{c}
 $-10.31$ \\
 $-3.63$
 \end{tabular}
 &
 \begin{tabular}{c}
 $[-11.96,-7.22]$ \\
 $[-5.10, -1.33]$
 \end{tabular}
 & 1.08\\ \\[-1.5ex]
 \hline\\[-1ex]
\end{tabular}
\end{table}

\begin{table}[htbp]
\caption{The same table as Table~\ref{tab:post_BOS} but for the LVK-C like loop distribution.}\label{tab:post_Model_C}
\begin{tabular}{ccccc}
 \\[-1ex]
 \multicolumn{5}{c}{\textbf{$\fng$-LVK-C models}}\\[0.5ex] 
 \hline
 Parameter & Bayes Est. &Maximum Post. & 68\% Credible Interval(s) & BF\\ [0.5ex] 
 \hline\hline\\[-1ex]
 \multicolumn{5}{c}{\textbf{String with $(N_c, N_k) = (2, 0)$}}\\[1ex]
 \begin{tabular}{c}
 $\log_{10}G\mu$ \\
 $\log_{10}\fng$
 \end{tabular}
 &
 \begin{tabular}{c}
 $-9.75 \pm 1.24$ \\
 $-1.30 \pm 1.32$
 \end{tabular}
 &
 \begin{tabular}{c}
 $-10.61$ \\
 $0^*$
 \end{tabular}
 &
 \begin{tabular}{c}
 $[-11.07, -9.36]$ \\
 $[-1.58, 0^*]$
 \end{tabular}
 & 1.00\\ \\[-1.5ex]
 \hline\\[-1ex]
 \multicolumn{5}{c}{\textbf{String with $(N_c, N_k) = (1, 100)$}}\\[1ex] 
 \begin{tabular}{c}
 $\log_{10}G\mu$ \\
 $\log_{10}\fng$
 \end{tabular}
 &
 \begin{tabular}{c}
 $-9.11 \pm 1.42$ \\
 $-2.21 \pm 1.41$
 \end{tabular}
 &
 \begin{tabular}{c}
 $-10.70$ \\
 $0^*$
 \end{tabular}
 &
 \begin{tabular}{c}
 $[-11.07,-10.11], [-9.54, -7.39]$ \\
 $[-3.70, -1.93], [-1.16, 0^*]$
 \end{tabular}
 & 0.46\\ \\[-1.5ex]
 \hline\\[-1ex]
 \multicolumn{5}{c}{\textbf{SMBHB + String with $(N_c, N_k) = (2, 0)$}}\\[1ex]
 \begin{tabular}{c}
 $\log_{10}G\mu$ \\
 $\log_{10}\fng$
 \end{tabular}
 &
 \begin{tabular}{c}
 $-10.25 \pm 1.76$ \\
 $-2.35 \pm 1.83$
 \end{tabular}
 &
 \begin{tabular}{c}
 $-10.66$ \\
 $0^*$
 \end{tabular}
 &
 \begin{tabular}{c}
 [$-11.3, -7.61$] \\
 $[-4.61, -3.03], [-1.67,0^*]$
 \end{tabular}
 & 1.32\\ \\[-1.5ex]
 \hline\\[-1ex]
 \multicolumn{5}{c}{\textbf{SMBHB + String with $(N_c, N_k) = (1, 100)$}}\\[1ex]
 \begin{tabular}{c}
 $\log_{10}G\mu$ \\
 $\log_{10}\fng$
 \end{tabular}
 &
 \begin{tabular}{c}
 $-10.32 \pm 2.00$ \\
 $-3.04 \pm 1.69$
 \end{tabular}
 &
 \begin{tabular}{c}
 $-10.99$ \\
 $-3.86$
 \end{tabular}
 &
 \begin{tabular}{c}
 [$-12.11, -7.54$] \\
 $[-5.23,-1.91], [-0.48, 0^*]$
 \end{tabular}
 & 1.14\\ \\[-1.5ex]
 \hline\\[-1ex]
\end{tabular}
\end{table}

\subsubsection{$\fng$-LVK-C models}
\begin{figure}[htbp] 
\centering
\includegraphics[clip,width=0.5\columnwidth]{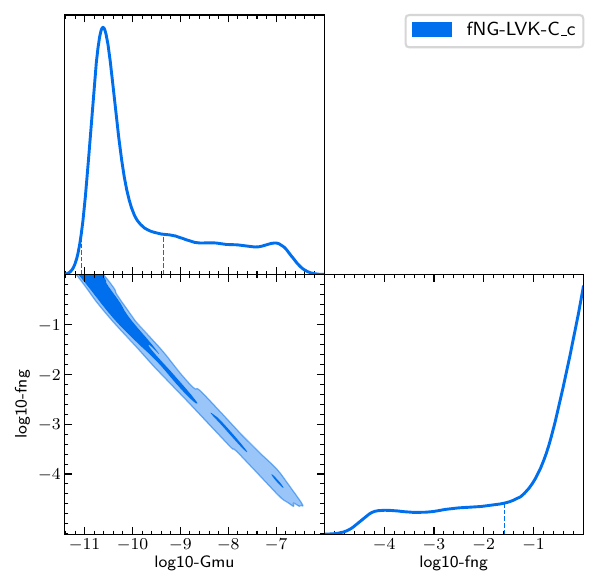}~
\includegraphics[clip,width=0.5\columnwidth]{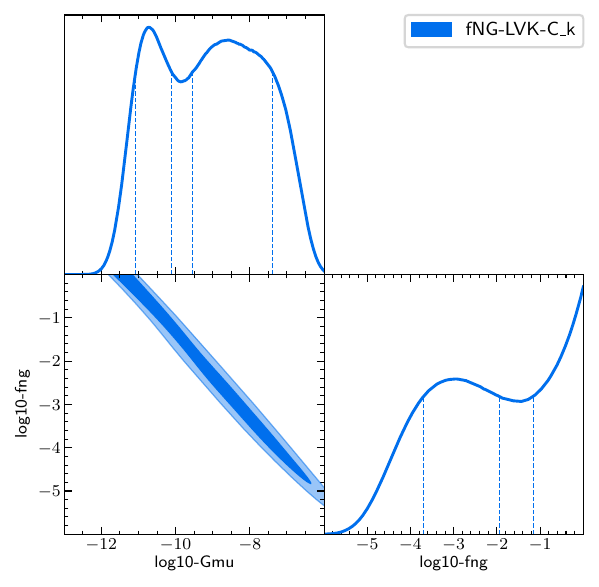}
  \caption{Posterior distribution of the string parameters for LVK-C-like models without SMBHB signal.}
    \label{fig:Model_C_fng}
\end{figure}

\begin{figure}[htbp] 
\centering
\includegraphics[clip,width=0.5\columnwidth]{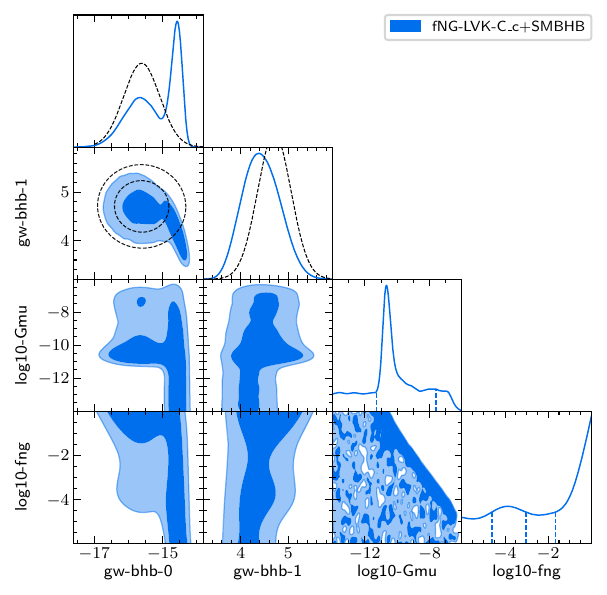}~
\includegraphics[clip,width=0.5\columnwidth]{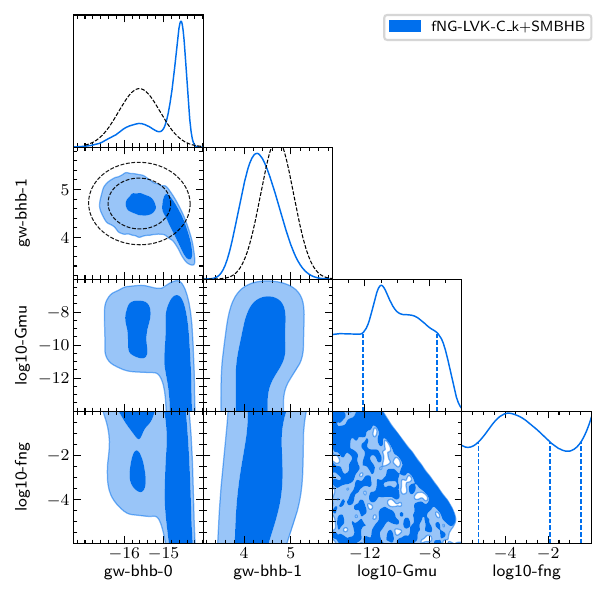}
  \caption{Posterior distribution of the string parameters for LVK-C-like models with SMBHB signal.}
    \label{fig:Model_C_fng_SMBH}
\end{figure}

Similarly to the previous sections, the posterior distributions of string (and SMBHB) parameters are shown in Fig.~\ref{fig:Model_C_fng} and Fig.~\ref{fig:Model_C_fng_SMBH}, respectively for the string signal only case and for the superposition case.
Let us note again that the C-1 scenario of LVK model C, which is referred in this analysis, connects the distribution of the matter-dominated era of the LRS model with the radiation-dominated era of the BOS model.
As this model predicts slightly larger SGWB amplitudes than those in LRS-like distribution, certain differences from Figs.~\ref{fig:LRS_fng} and~\ref{fig:LRS_fng_SMBH} are observed.
Indeed, 3 out of 4 cases, Model-C-like distribution yields higher Bayes factor $(\mathcal{B}_{10} \gtrsim 1)$ than those in LRS-like case.

Nevertheless, the overall behaviour is quite similar to those for the LRS-like distribution and they are subjected to the existing limits in the similar way.
Therefore, kink-dominated scenario (right panels of Figs.~\ref{fig:Model_C_fng} and~\ref{fig:Model_C_fng_SMBH}) again seems to be more compatible with the simulation bound $\fng \lesssim 0.1$.
This is quite reasonable since the loops created in the matter dominated era gives the dominant contribution in the PTA band.
Therefore, we expect that when C-2 scenario is considered as the reference model, the results should resemble those for the BOS-like distribution.

\section{Discussion}\label{sec:discussion}
In this work, we performed Bayesian inference analyses with NANOGrav 15yrs observational data to derive implications for local cosmic strings, taking into account the largest theoretical uncertainty, which is the lifetime of cosmic string loops. 
In contrast to the conventional NG description, field theoretic simulations in the AH model indicates that the primary decay channel of string loops is classical radiation of massive particle. Following the previous study~\cite{Hindmarsh:2022awe}, we therefore assume that a fraction $\fng$ of the AH string loops are NG-like and assume that the loop distribution inferred from NG simulations~\cite{Lorenz:2010sm,Blanco-Pillado:2013qja}.
This parametrization, which takes into account the observation that not all the string loops radiate GWs in the AH model, leads to a SGWB which is a factor $\fng$ smaller than the NG prediction.
Therefore, the observational GW data puts constraints on the combination of the string tension $G\mu$ and $\fng$, not on $G\mu$ alone. 

In quantifying the SGWB signal from long-lived AH string loops, here we took three representative models for NG loop distribution~\cite{Lorenz:2010sm, Blanco-Pillado:2013qja, Auclair:2019zoz}.
These models are only approximations since there are no long-lived loops observed in simulations of networks the AH model. 
Therefore, the interpretation of $\fng$ as a fraction of long-lived loops and the constraints on $(G\mu, \fng)$ derived in the analyses physically make sense only if the model distribution is close to the actual loop distribution.
In this sense, the adoption of the three models is a case-by-case examination of three distinct possibilities, each assuming that the real distribution of long-lived loops matches that of a particular NG model.
Further study of possible NG-like loop production is indispensable for the refinement of SGWB constraint on the AH model: the current status is only that $\fng \lesssim 0.1$ and no information on ${\sf n}(l,t)$.

Our analyses were based on the wrapper \texttt{PTArcade}~\cite{Mitridate_2023,Mitridate:2023oar} and conducted in the similar way to the NANOGrav study of new physics \cite{NANOGrav:2023hvm}. 
Note that while the BOS model was analysed as a NG loop distribution model and subjected to the model selection analysis by NANOGrav~\cite{NANOGrav:2023hvm}, the LRS and LVK-Model-C were not (see Ref.~\cite{EPTA:2023xxk} for the posterior for the LRS model derived from EPTA data).
As summarized in Sec.~\ref{sec:data_analysis}, we performed Bayesian inference and model selection analysis for all three models assuming only the signal from strings and also assuming the signal from SMBHB. We also investigated the different loop power spectra for each model.
In all cases our posterior for the AH model extends to a higher value of $G\mu$ (with a decrease in the value of $\fng$), compared to the pure NG model results in Refs.~\cite{NANOGrav:2023hvm,EPTA:2023xxk}.
Phenomenologically, this suggests that symmetry breaking in the AH model is still possible on relatively high energy scales.
It is also worth mentioning that for all the loop distribution models, the posterior of SMBHB parameters become more consistent with the theory-motivated prior used in the NANOGrav paper~\cite{NANOGrav:2023gor}, when the string signal is superimposed to the SMBHB signal. 
When comparing each model, the LRS-like and model-C-like distribution overall had larger Bayes factors (against the SMBHB signals) than the BOS-like distribution. This is due to the fact that for a wider range of $G\mu$ values, loops of the LRS model and LVK model-C produce a blue-tilted SGWB, which is favored by the data.
Note again that the differences in our results for different reference NG loop models should be regarded as an estimate of the uncertainty in the AH loop distribution.
%We should, however, note that the differences in our results for different reference NG loop models should be regarded as an estimate of the uncertainty in the AH loop distribution.

After deriving the posterior distributions, we compared them to the existing bounds such as $G\mu \lesssim 10^{-7}$ from CMB~\cite{Planck:2013oqw,Planck:2015fie} and $\fng \lesssim 0.1$ derived from the AH loop simulation~\cite{Hindmarsh:2021mnl}.
For the BOS-like distribution (without SMBHB signal), we found that the posterior might be in tension with those constraints. The loop simulation bound also seems to constrain the $(N_c, N_k) = (2, 0)$ scenario both for the LRS-like and the Model-C-like distributions.
Therefore, we may conclude that the AH model could explain the data by itself if the NG-like loops follow LRS model or Model-C with kink-dominated loop spectrum.
This may cast a doubt on the AH model as the origin of the SGWB becoming visible at the PTAs, given that the validity of LRS model has been recently questioned in terms of energy transfer (see {\it e.g.} Ref.~\cite{Blanco-Pillado:2019vcs}).
Otherwise, they should be sub-dominant over the other plausible signals such as the SMBHB signal.
In this case, SGWB observations becomes not so constraining for the AH model due to the degeneracy of $G\mu$ and $\fng$ in determining the signal amplitude. To further investigate the model, one should combine the constraint from the particle radiation as discussed in Refs.~\cite{Hindmarsh:2013jha, SantanaMota:2014xpw, Hindmarsh:2022awe}, which can place the upper bounds on $G\mu$ depending on the interaction between the AH sector and the other sectors.

\section*{Acknowledgments}
The authors would like to thank Andrea Mitridate for his kind introduction to \texttt{PTArcade} and thank Kai Schmitz and Tobias Schröder for the fruitful discussion.
JK (ORCID ID 0000-0003-3126-5100) is supported by the JSPS Overseas Research Fellowships.
MH (ORCID ID 0000-0002-9307-437X) acknowledges support from the Academy of Finland grant no.~333609. 

\appendix
\section{Test analysis with the pure NG model}\label{app:exact_BOS}
\begin{figure}[htbp] 
\centering
\includegraphics[clip,width=0.4\columnwidth]{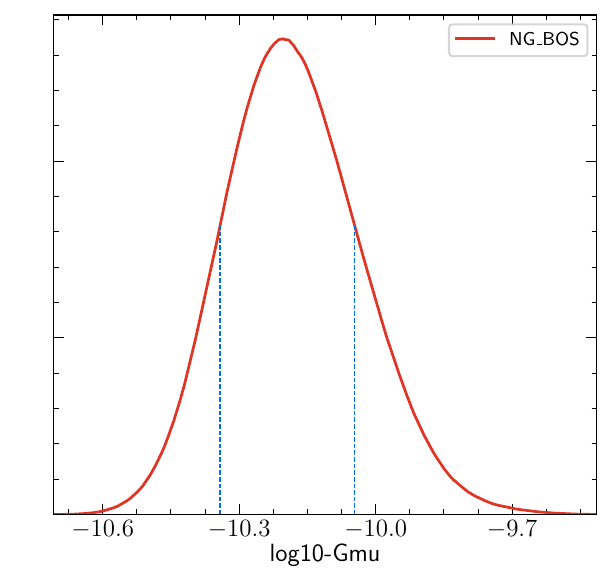}~
\includegraphics[clip,width=0.5\columnwidth]{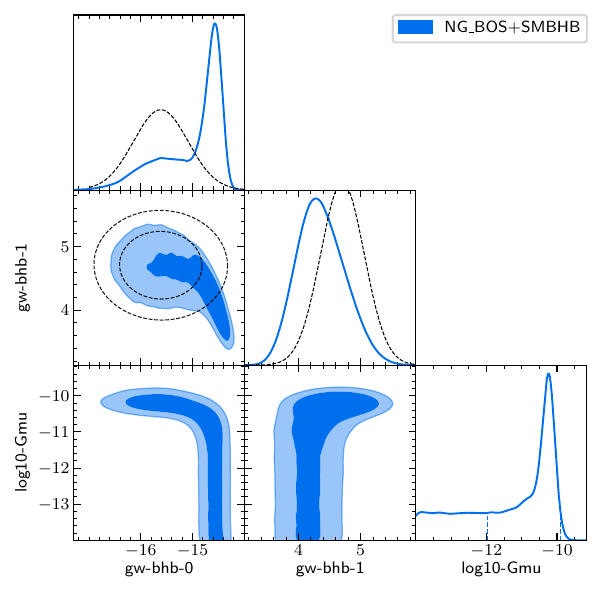}
  \caption{Posterior distribution of the model parameters for the BOS loop distribution. The left panel is for string signal only and the right one for the superposition.}
    \label{fig:NG_BOS}
\end{figure}

Here we summarize the result of our test analysis, where the pure BOS model ($\fng = 1$) is assumed. The outputs of analysis are summarized in Table.~\ref{tab:post_exact_BOS} and Fig.~\ref{fig:NG_BOS}, which can be compared with the result of ``STABLE-N'' model in the NANOGrav paper (see Table~4 in Ref.~\cite{NANOGrav:2023hvm}).
Although we did not generate the same number of chains as Ref.~\cite{NANOGrav:2023hvm}, we confirmed that our results (posterior distribution in Fig.~\ref{fig:NG_BOS} and the quantities in Table.~\ref{tab:post_exact_BOS}) agreed well with their results. %\footnote{Although the $68\%$ credible interval is written as $[-11.99,-9.90]$ for STABLE-N + SMBHB in Ref.~\cite{NANOGrav:2023hvm}, we confirmed that it becomes $[-14^*,-9.90]$ when we subject their public chain to {\texttt{PTArcade}}. \MHc{What are we confirming?  Seems like we are differing from them.}\JKc{Right. I suspect that they mistakenly wrote an incorrect value on their paper. What I want to say here is that at least we are consistent.}}.
The agreement is a useful check of the results of our analysis. 

\begin{table}[htbp]
\caption{Summary of the results for the pure BOS distribution. We confirm that our result well agrees with those from the NANOGrav collaboration analysis~\cite{NANOGrav:2023hvm}.}\label{tab:post_exact_BOS}
\begin{tabular}{ccccc}
 \\[-1ex]
 \multicolumn{5}{c}{\textbf{BOS distribution (``STABLE-N'' model)}}\\[0.5ex] 
 \hline
 Parameter & Bayes Estimator & Maximum Posterior & 68\% Credible Interval & Bayes Factor\\ [0.5ex] 
 \hline\hline\\[-1ex]
 \multicolumn{5}{c}{\textbf{String with smoothed loop model}}\\[1ex] 
 $\log_{10}G\mu$
 &
 $-10.18 \pm 0.15$ 
 &
 $-10.20$
 &
 $[-10.34, -10.04]$ 
 & 0.32\\ \\[-1.5ex]
 \hline\\[-1ex]
 \multicolumn{5}{c}{\textbf{SMBHB + String with smoothed loop model}}\\[1ex] 
 $\log_{10}G\mu$
 &
 $-11.32 \pm 1.23$
 &
 $-10.24$ 
 &
 $[-11.97, -9.90]$ 
 & 0.84\\ \\[-1.5ex]
 \hline\\[-1ex]
\end{tabular}
\end{table}

\section{Changing prior for the SMBHB signal}\label{app:prior}
Given the uncertainty in the theoretical prediction for the SMBHB signal, here we consider a different prior choice for its parameters and discuss its effect on the results. 
For simplicity, let us assume a log-uniform prior distribution for both parameters $(\log_{10}A_{\rm BHB}, \gamma_{\rm BHB})$ respectively as $[-18, -12]$ and $[0, 7]$, which can be easily specified in \texttt{PTArcade}. 

\begin{figure}[htbp] 
\centering
\includegraphics[clip,width=0.5\columnwidth]{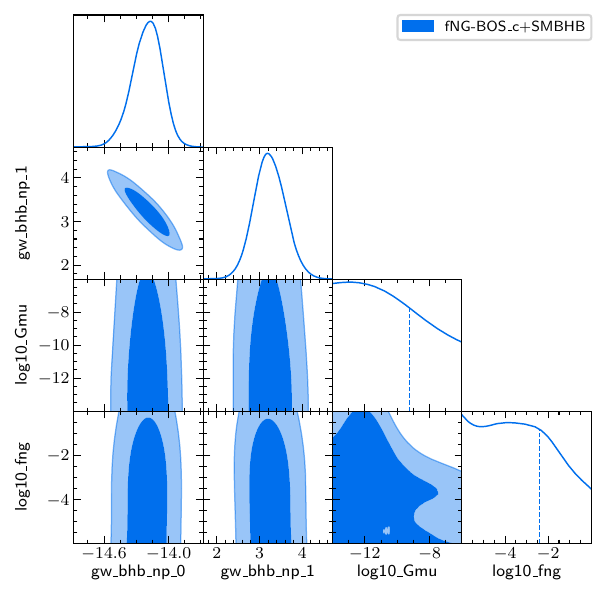}~
\includegraphics[clip,width=0.5\columnwidth]{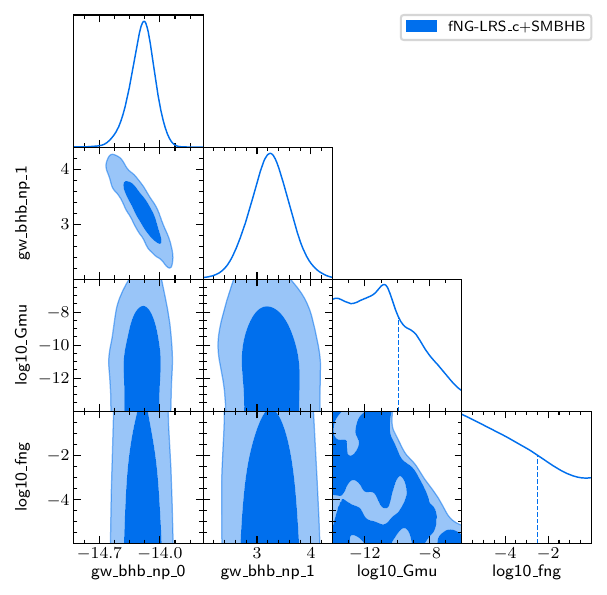}
  \caption{Posterior distribution of the model parameters with flat log-uniform prior assumed for power-law signal. 
  The left and right panels are for the same model as the left panel of Fig.~\ref{fig:BOS_fng_SMBH} and that of Fig.~\ref{fig:LRS_fng_SMBH}, respectively.}
    \label{fig:diff_LRS_fng_SMBH}
\end{figure}

Assuming the BOS-like model with smooth loop power and the LRS-like model with $(N_c, N_k) = (2,0)$ for string signal, we respectively consider the superposition of the string signal and the power-law (SMBH) signal.
In Fig.~\ref{fig:diff_LRS_fng_SMBH}, the posterior distribution of the model parameters are shown. Here, {\texttt gw\_bhb\_np\_0} and {\texttt gw\_bhb\_np\_1} represent $A_{\rm BHB}$ and $\gamma_{\rm BHB}$ respectively.
Especially for LRS-like model, one can see the clear differences from the left panel of Fig.~\ref{fig:LRS_fng_SMBH} where the theoretical prior is assumed for SMBHB signal.
By comparing the 2D posterior of power-law parameters, we can no longer observe the string signal dominated branch in the present case.
This is because the prior of the SMBHB signal is assumed to be flat over sufficiently broad region, allowing the data to be fitted with an ideal power-law signal.
Consequently, the marginalized distribution of the string parameters also significantly differs.
Although this is an extreme example of the prior dependence, we should carefully follow future results on the SMBHB signal prediction since it could affect the constraint on the string parameters.

\if0
\appendix
\section{Varying the number of kinks $N_{\rm k}$}~\label{app:kink_num}
given the uncertainty on the initial number of kinks at the loop production, one might consider...  
As discussed in Refs.~\cite{EPTA:2023hof, EPTA:2023xxk}, however,...
\fi

\bibliographystyle{JHEP}
\bibliography{main}

\end{document}